\pdfoutput=1

\documentclass[twocolumn,english,aps,superscriptaddress,prb]{revtex4-2}

\usepackage{babel}
\usepackage{amsmath}
\usepackage{amssymb}
\usepackage{graphicx}

\newcommand{\appropto}{\mathrel{\vcenter{
  \offinterlineskip\halign{\hfil$##$\cr
    \propto\cr\noalign{\kern2pt}\sim\cr\noalign{\kern-2pt}}}}}

\begin{document}

\title{From SU(2)$_5$ to SU(2)$_3$ Wess-Zumino-Witten transitions in a frustrated spin-5/2 chain}

\author{Natalia Chepiga}
\affiliation{Kavli Institute of Nanoscience, Delft University of Technology, Lorentzweg 1, 2628 CJ Delft, The Netherlands}
\author{Ian Affleck}
\affiliation{Stewart Blusson Quantum Matter Institute, Department of Physics and Astronomy, University of British Columbia, Vancouver, BC, Canada V6T 1Z1}
 \author{Fr\'ed\'eric Mila}
 \affiliation{Institute of Physics, Ecole Polytechnique F\'ed\'erale de Lausanne (EPFL), CH-1015 Lausanne, Switzerland}

\date{\today}
\begin{abstract}
We investigate the properties of a frustrated spin-5/2 chain with next-nearest neighbor two and three-site interactions, with two questions in mind: the nature of the transition into the dimerized phase induced by the three-site interaction, and the possible presence of a critical floating phase at intermediate values of the next-nearest neighbor interaction. We provide strong evidence that the continuous transition into the dimerized phase, which has been found to be generically in the Wess-Zumino-Witten SU(2)$_{2S}$ universality class up to spin $S=2$, is SU(2)$_5$ only at two isolated points of the phase diagram, and that it is SU(2)$_3$ in between, in agreement with the presence of two relevant operators allowed by symmetry for SU(2)$_5$, and with the conservation of the parity of the level index along the renormalization flow between SU(2)$_k$ theories with different values of $k$. We also find that the dimerization induced by the next-nearest neighbor interaction is a three step process, with first a small partially dimerized phase followed by a broad critical floating phase with incommensurate correlations before the fully dimerized phase is reached. Implications for the iron oxide ${\mathrm{Bi}}_{3}{\mathrm{FeMo}}_{2}{\mathrm{O}}_{12}$ are briefly discussed.
\end{abstract}

\maketitle

\section{Introduction}

Antiferromagnetic Heisenberg spin chains have attracted a lot of attention over the years. Competing interactions induce frustration and are known to lead to new phases and quantum phase transitions. 
For example, the $J_1-J_2$ spin-1/2 chain undergoes a Kosterlitz-Thouless transition\cite{Kosterlitz} into a spontaneously dimerized phase\cite{MajumdarGhosh} when the ratio of the next-nearest neighbor interaction to the nearest neighbor one $J_2/J_1>0.24112$\cite{okamoto}. The picture is radically different for spin-1 chain where both phases realized by $J_1-J_2$ model are gapped and non-dimerized. For small $J_2$ the chain is in the topologically non-trivial Haldane phase\cite{Haldane}, while at large values of $J_2$ the ground-state corresponds to two inter-twinned Haldane chains\cite{kolezhuk_connectivity,kolezhuk_prl,kolezhuk_prb}, and a first order transition separates the two phases.

Recently, it has been shown that the three-site interaction $J_3 [({\bf S}_{i-1}\cdot {\bf S}_i)({\bf S}_{i}\cdot {\bf S}_{i+1})+\mathrm{h.c.}]$ induces a fully dimerized state in spin-$S$ chains\cite{michaud1}. In fact, at $J_3=J_1/[4S(S+1)-2]$ the fully dimerized state is an exact ground-state\cite{michaud1}. 
For spin-1/2 the $J_3$ term reduces to the next-nearest-neighbor interaction, while for larger spin the combination of these two types of frustration lead to a very rich phase diagram\cite{j1j2j3_short,j1j2j3_long,spin_32paper}.
The exactly dimerized ground state found for the $J_3$ interaction has been shown to remain an exact eigenstate along the line\cite{wang}: 
\begin{equation}
  \frac{J_3}{J_1-2J_2}=\frac{1}{4S(S+1)-2},
  \label{eq:exact}
\end{equation} 
but it only remains the ground state for not too small $J_3$.

The numerical investigation of these models has been primarily focused on the nature of the quantum phase transitions into the dimerized phase and has already come up with a number of surprising results.
In particular, the quantum phase transition between the fully-dimerized phase and the next-nearest neighbor (NNN-) Haldane phase in the spin-1 chain has been shown to be continuous in the Ising universality class while singlet-triplet gap never closes across the transition\cite{j1j2j3_short}. The transitions between the non-dimerized phase at small $J_2$ and the fully-dimerized phase were identified to be in the the Wess-Zumino-Witten (WZW) SU(2)$_{k=2S}$ universality classes at least for $1/2\leq S\leq 2$\cite{michaud2}, and interesting result for the realization of the WZW SU(2)$_{2S}$ universality classes that are otherwise better known to appear in integrable spin-$S$ chains with a microscopic Hamiltonian given by a polynomial of degree 2$S$ in $({\bf S_i}\cdot {\bf S_{i+1}})$\cite{kulish,takhtajan,babujian}. In spin-1 and spin-3/2 chains, due to the presence of marginal operators, these continuous WZW SU(2)$_{2S}$ transitions turn into first order transitions upon increasing the next-nearest-neighbor coupling $J_2$. This result came as a surprise for the spin-3/2 chain because it implies the existence of a counter-intuitive first-order transition between the critical SU(2)$_1$\cite{affleck_haldane} and the fully-dimerized phases. More importantly for our present purpose, the straightforward generalization of WZW SU(2)$_{k=2S}$ to larger spins contradicts field theory arguments\cite{lecheminant}, and the nature of the dimerization transitions for $S\geq 5/2$ remains an open problem.

Previous investigations of the $J_1-J_2-J_3$ model also revealed that for a large portion of the phase diagram spin-spin correlations are incommensurate\cite{kolezhuk_prl,kolezhuk_prb,roth,j1j2j3_long,spin_32paper}. In the spin-1 chain, where all three phases - Haldane, NNN-Haldane and dimerized - are gapped, incommensurate correlations can only be seen as short-range order. We argued\cite{spin_32paper} previously that, for the spin-1 chain, these incommensurate correlations can be interpreted as twists of the valence-bond configuration with an average distance between the twists that is related to the wave-vector of the incommensurate phase, and that can be tuned continuously with $J_2$ and $J_3$. When such a periodically twisted Haldane chain is coupled with a critical spin-1/2 chain, as in the spin-3/2 chain at $J_2\sim J_1$, the short-range incommensurability turns into quasi-long-range order, resulting in a floating phase\cite{spin_32paper}. One can expect this feature to be generic for all zig-zag chains with half-integer spins.

In the present paper we study the combined effect of next-nearest-neighbor and three-site interactions in the
spin-5/2 chain. The model is defined by the following $J_1-J_2-J_3$ Hamiltonian:
\begin{multline}
  H=J_1\sum_i {\bf S}_i\cdot{\bf S}_{i+1}+J_2\sum_i {\bf S}_{i-1}\cdot{\bf S}_{i+1}\\
  +J_3\sum_i\left[({\bf S}_{i-1}\cdot {\bf S}_i)({\bf S}_i\cdot {\bf S}_{i+1})+\mathrm{ h.c.}\right].
  \label{eq:j1j2j3s}
\end{multline}
We focus on $J_1,J_2,J_3>0$, and without loss of generality we fix $J_1=1$. At $J_2=J_3=0$, the system is critical in the WZW SU(2)$_1$ universality class\cite{affleck_haldane}. To explore the rest of the phase diagram we rely on numerical simulations performed with a state-of-the-art density matrix renormalization group (DMRG) algorithm\cite{dmrg1,dmrg2,dmrg3,dmrg4}. Throughout the paper, unless explicitly stated otherwise, we use a chain with an even number of sites and open boundary conditions. In two-site DMRG we typically keep up to 1800 states, perform 6 sweeps and discard singular values smaller than $10^{-8}$.

The rest of the paper is organized as follows. In Sec.\ref{sec:phasediag} we discuss the main properties of the phase diagram. In Sec.\ref{sec:wzw} we discuss at length the nature of the transition between the critical and the fully-dimerized phase. In Sec.\ref{sec:float}, we focus on the floating phase - critical phase with quasi-long-range incommensurate order. Finally, we discuss the implications of our results and put them in perspective in Sec.\ref{sec:discussion}.

\section{\label{sec:phasediag}Phase diagram}

\begin{figure}[t!]
\centering 
\includegraphics[width=0.5\textwidth]{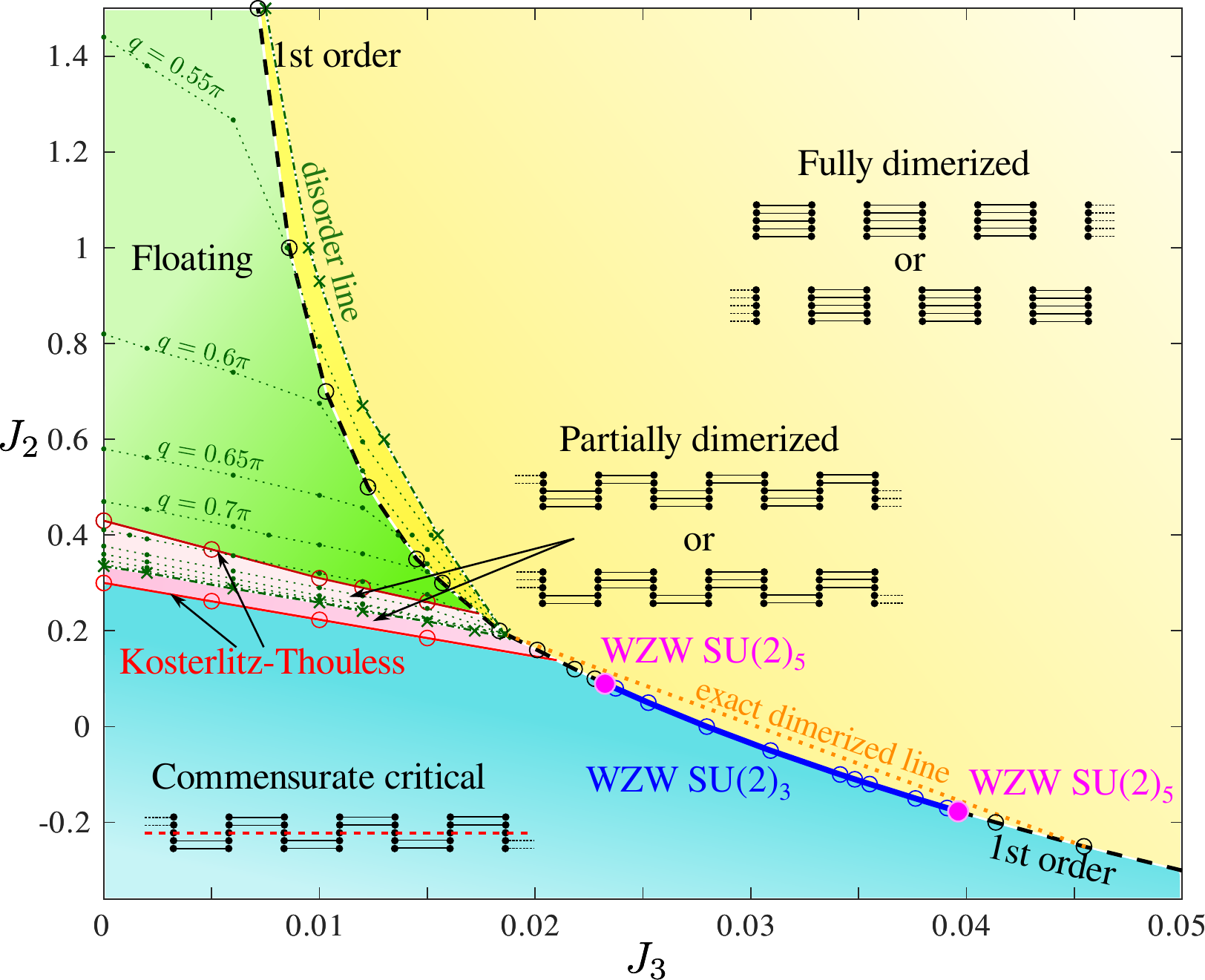}
\caption{Phase diagram of the $S=5/2$ chain with next-nearest-neighbor $J_2$ and three-site interactions $J_3$. Both partially and fully dimerized phases are gapped and spontaneously break the translation symmetry.  Both the commensurate critical   and the incommensurate critical (floating) phases are characterized by gapless spectra and algebraically decaying correlations. The fully dimerized phase is separated from the commensurate critical phase by a continuous WZW SU(2)$_3$ transition along the blue solid line that terminates at the two end points (magenta circles) located at $J_2\approx 0.09$ and $J_3\approx 0.0232499$ and $J_2\approx -0.177$ and $J_3\approx 0.039628$, both in the WZW SU(2)$_5$ universality class. The partially dimerized phase is separated from both the floating and critical phases by Kosterlitz-Thouless critical lines (light and dark red). The transition between the partially and fully dimerized phases is always first order (thick dashed line). In the parameter range of the figure, the transition between the floating phase and the fully dimerized phase is always first order. The ground-state is exactly dimerized along the orange dotted line given by Eq.(\ref{eq:exact}). The dashed green lines inside the fully and partially dimerized phases stand for the disorder lines where the correlations become incommensurate. The dotted green lines and gray dash-dotted lines are disorder lines. The sketches are visualizations of the corresponding phases in terms of valence bond singlets (VBS): dots state for spin-1/2, black lines for VBS singlets and the red dashed line for a VBS singlet resonating between two nearest-neighbor bonds.  }
\label{fig:phasediag}
\end{figure}

The ground-state phase-diagram obtained with DMRG is shown in Fig.\ref{fig:phasediag}. It consists of four phases: an SU(2)$_1$ commensurate critical phase at small $J_2$ and $J_3$, a fully-dimerized phase at large $J_3$ and/or at very large $J_2$, a partially dimerized phase at $J_2\simeq 0.3-0.4$ and small $J_3$, and an extended critical floating phase 
with incommensurate correlations that starts above the partially dimerized phase and extends up to large values of $J_2$.

The system exhibits a very rich critical behavior. Apart from the already mentioned commensurate and incommensurate (floating) critical phases, the system undergoes a continuous WZW SU(2)$_3$ phase transition at small values of $J_2$ (both positive and negative). This continuous transition terminates on both sides at a critical end point in the WZW SU(2)$_5$ universality class, beyond which the transition between the critical and fully dimerized phases becomes first order.
A similar first-order transition between a critical and a gapped (fully-dimerized) phase has been recently reported for the spin-3/2 chain\cite{spin_32paper}. To get an intuition on this rather exotic transition, one can think of a level crossing between a two-fold degenerate ground-state on the dimerized side and a continuum of low-lying states on the critical side.

On the phase diagram we can distinguish two different dimerized phases with partial and full dimerization. The partially dimerized phase is separated from both the commensurate and the incommensurate critical phases by a Kosterlitz-Thouless transition. In terms of valence bond singlets (VBS) this phase can be visualized as five singlets located on every other nearest-neighbor bond. Based on the numerical value of the dimerization inside the partially dimerized phase we visualize it as alternating double and triple VBS singlets. We want to stress, however, that the edge states that might appear in this phase are not protected, and that there are other VBS representations topologically equivalent to those shown in Fig.\ref{fig:phasediag} that lead to partial dimerization.
 The incommensurate correlations typical of the floating phase appear as short-range incommensurate correlations inside the two gapped dimerized phases surrounding the floating phase. In Fig.\ref{fig:phasediag} we show the location of these disorder lines where incommensurate values of the wave-vector $q$ start to dominate. We also mark the equal-$q$ lines along which the wave-vector $q$ takes constant values. 

We also provide numerical evidence that, at large $J_2$, the floating phase disappears to give way to the fully-dimerized phase, similar to the spin-3/2 case. Nevertheless, the floating phase is remarkably extended and continues along $J_3=0$ up at least $J_2\sim 6$. 

Below we discuss the various critical regimes in details. 

\section{Wess-Zumino-Witten critical line and the two end points}
\label{sec:wzw}

We start our discussion with the most intriguing part of the phase diagram - the continuous transition between the commensurate critical phase and the fully dimerized phase.
Previous studies of the $J_1-J_3$ model for spin $S\leq 2$ \cite{michaud1,michaud2} identified the transitions to the fully dimerized phase to be in the WZW SU(2)$_{2S}$ universality class. More recently, it has been shown that in the presence of next-nearest-neighbor interactions $J_2$, these critical points are part of an extended WZW SU(2)$_{2S}$ critical lines that, for both spin-1 and spin-3/2, eventually turns into a first order transition due to the presence of a marginal operator. 

However, the naive generalization of this conclusion to spins $s\geq 5/2$ would violate a field theory argument raised in Ref.\cite{lecheminant}. Let us briefly recap the argument here. In WZW SU(2)$_5$ critical theory there are two relevant operators with dimensions $x=2/7$ and $x=6/7$ allowed by symmetry. By fine tuning two external parameters (in our case $J_2$ and $J_3$) one can eventually drive a system into an isolated WZW SU(2)$_5$ critical point.  However, by tuning only one parameter, for instance by scanning along a critical line, it is not possible to get rid of both relevant operators. Thus the scenario where the WZW SU(2)$_5$ critical theory would be realized along a line in a two-dimensional parameter space can be discarded. 

This leads to a very natural question: if the entire continuous phase transition that we observe between the commensurate critical and the fully dimerized phases cannot be in WZW SU(2)$_5$ universality class, then what would be the underlying critical theory of this transition?  It is worth noticing that the entire critical line including its end points is adjacent to the commensurate critical phase described by the WZW SU(2)$_1$ theory. Now, according to Furuya and Oshikawa, \cite{PhysRevLett.118.021601} in the presence of both SU(2) and a discrete  $\mathbb{Z}_2$ symmetry of the WZW theory, which corresponds to the translation symmetry for an antiferromagnetic spin chain, a renormalization-group flow is only possible between WZW SU(2)$_k$ theories if the parity of the level index $k$ does not change. This suggests that there are only two possible candidates to which the WZW SU(2)$_{k=5}$ can renormalize: either to SU(2)$_{k=1}$, the critical theory in the bulk of the critical phase, or to SU(2)$_{k=3}$, the critical theory that describes e.g. the transition into the dimerized phase of the spin-3/2 chain\cite{spin_32paper}.

\begin{figure}[t!]
\centering 
\includegraphics[width=0.5\textwidth]{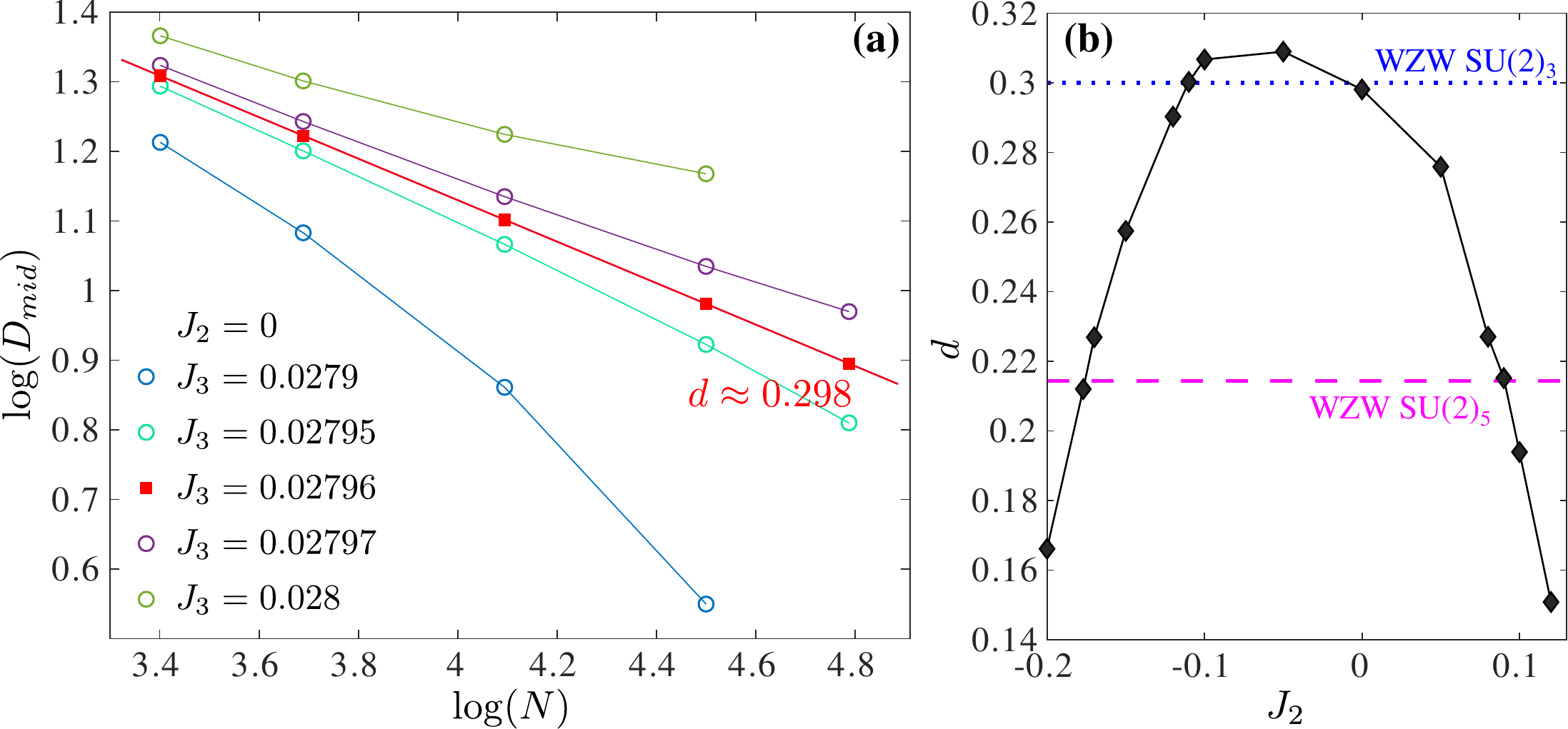}
\caption{(a) Example of finite-size scaling of the middle-chain dimerization for $J_2=0$. The critical point is associated with the separatrix at $J_3\approx0.02796$, and the slope gives an effective scaling dimension $d_{eff}\approx0.298$. (b) Resulting value of the effective scaling dimension $d_{eff}$ as a function of $J_2$ along the transition between the commensurate critical and the fully dimerized phases. We associate the end points with the crossing points of the resulting curve and the horizontal line $d_{k=5}=3/14$ (dashed magenta). The dashed blue line stands for $d_{k=3}=3/10$. }
\label{fig:dimeriscaling}
\end{figure}

In order to explore the nature of the critical line we first have to locate it on the phase diagram. For this, we look at the finite-size scaling of the middle chain dimerization $D_{mid}=|\langle {\bf S}_{N/2-1}\cdot {\bf S}_{N/2} \rangle-\langle {\bf S}_{N/2}\cdot {\bf S}_{N/2+1} \rangle|$. Fig.\ref{fig:dimeriscaling}(a) shows an example of such a scaling for $J_2=0$ plotted in a log-log scale. Convex curves lead to a finite dimerization in the thermodynamic limit and are associated with the dimerized phase, while concave curves point towards the non-dimerized phase (later we will come back to this point with an argument on why close to the transition the scaling with the size forms a concave curve, unlike in the middle of the commensurate critical phase, where the scaling is linear). Therefore, we associate the transition point with the separatrix, the slope of which gives an estimate of the scaling dimension $d$. Note that this method has an excellent resolution and allows one to locate the critical point $J_3^c$ with an accuracy of $O(10^{-6})$.

The scaling dimension $d$ can be expressed as a combination of two standard critical exponents as $d=\beta/\nu$, where $\beta$ controls how the dimerization vanishes and $\nu$ how the correlation length diverges upon approaching the phase transition from the ordered (read fully-dimerized) phase. In general, for WZW SU(2)$_k$ these critical exponents are controlled by the level index $k$:
\begin{equation}
  \nu=\frac{2+k}{2k},\ \ \ \beta=\frac{3}{4k},
\end{equation}
leading to the scaling dimension:
\begin{equation}
d=\frac{3}{2(2+k)}.
\end{equation}

By keeping track of the extracted scaling dimension $d$ along the transition also for negative (ferromagnetic) next-nearest-neighbor coupling $J_2$ we obtained the results presented in Fig.\ref{fig:dimeriscaling}(b) and came to the following conclusions: {\it i)} the curve never approaches the value $d=1/2$, excluding the WZW SU(2)$_1$ from the possible candidates to describe the critical line;  {\it ii)} for an extended interval of $J_2$ the scaling dimension takes values around $d\approx 0.3$, in agreement with our second candidate WZW SU(2)$_3$; {\it iii)} if somewhere along the transition line there are isolated points in the WZW SU(2)$_5$ universality class, they have to be located at points where the scaling dimension takes the values $d=3/14\approx0.214$. According to Fig.\ref{fig:dimeriscaling}(b), this happens twice at $J_2\approx-0.177$ and at $J_2\approx 0.09$. At this stage the last statement is a conjecture that we are going to test.

\begin{widetext}

\begin{figure}[t!]
\centering 
\includegraphics[width=0.9\textwidth]{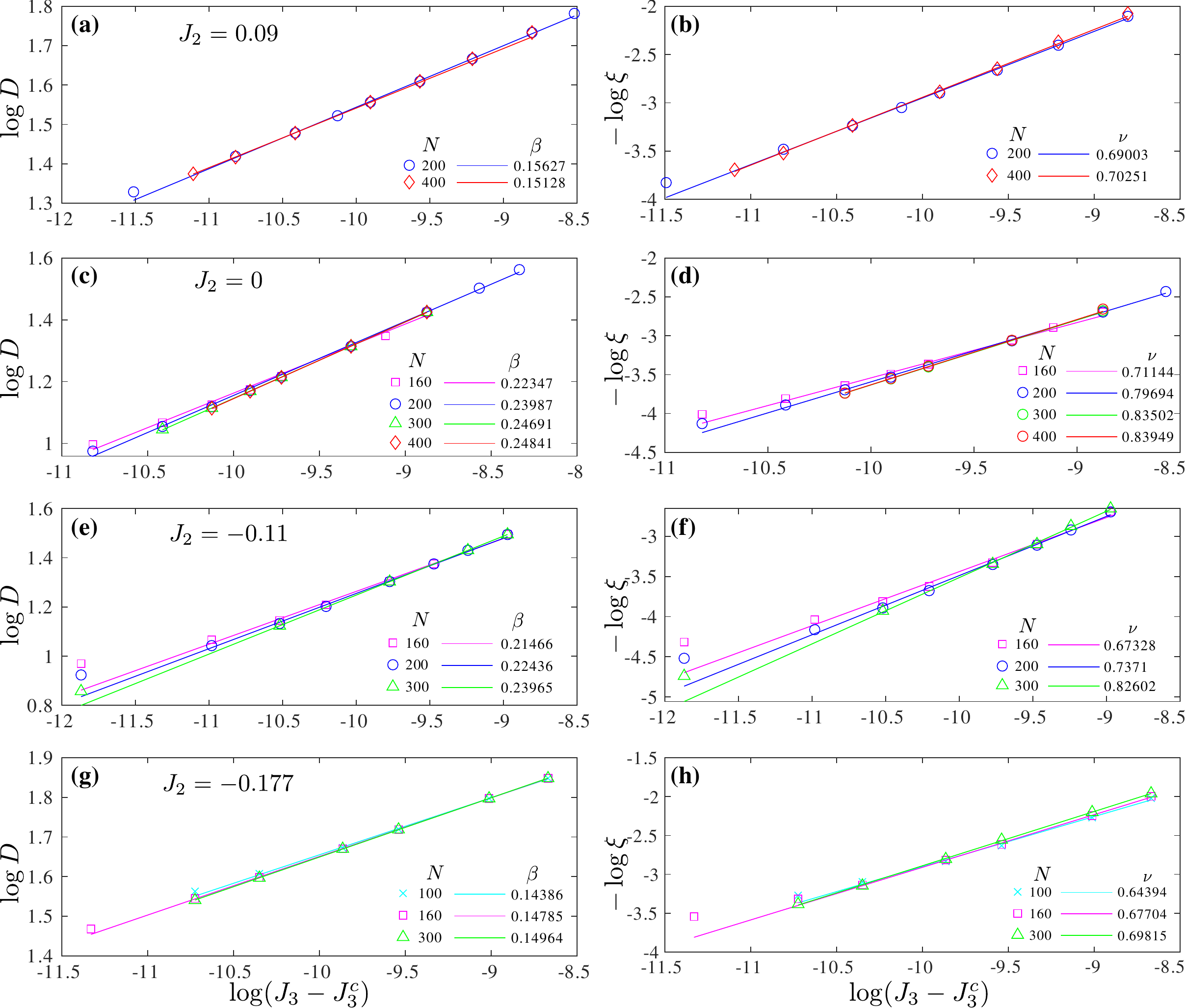}
\caption{ Scaling of the middle chain dimerization $D$ (left panels) and of the correlation length $\xi$ upon approaching the phase transition from the fully dimerized phase to the commensurate critical phase along four horizontal cuts. (a-b) Cut through the upper end point ($J_2=0.09$); the extracted critical exponents are in excellent agreement with the CFT predictions for WZW SU(2)$_5$, $\beta=3/20$ and $\nu=7/10$. (c-f) Cuts through the continuous transition at (c-d) $J_2=0$ and (e-f) $J_2=-0.11$. In both cases, the critical exponents are in good agreement with the CFT predictions for WZW SU(2)$_3$, $\beta=1/4$ and $\nu=5/6\approx 0.833$. One can notice that the finite-size effect become stronger for negative values of $J_2$. (g-h) Cut through the lower end point at $J_2\approx-0.177$. The results are again in a good agreement with the values of WZW SU(2)$_5$ critical exponents.}
\label{fig:end_points}
\end{figure}

\end{widetext}

To check our hypothesis regarding the nature of the critical line, we have extracted the critical exponents $\beta$ and $\nu$ independently. In Fig.\ref{fig:end_points} we present the results for four different cuts: at $J_2\approx 0.09$ (a-b) and at $J_2\approx-0.177$ (g-h), where the WZW SU(2)$_5$ universality class can appear, and along two arbitrary cuts through the critical line where $d\approx0.3$ - we have chosen $J_2=0$ (c-d) and $J_2=-0.11$ (e-f). For each cut we fix the location of the critical point to the value obtained form the finite-size scaling analysis of the dimerization ($J_3^c\approx 0.023249$ for $J_2=0.09$, $J_3^c\approx 0.02796$  for $J_2=0$, $J_3^c\approx 0.034823$  for $J_2=-0.11$, and $J_3^c\approx 0.039628$  for $J_2=-0.177$), and we perform simulations on chains of different lengths to track the finite-size effects. The values of the exponents extracted in this way can be found in the panels of Fig.\ref{fig:end_points}.

For both end points at $J_2\approx0.09$ and $J_2\approx-0.177$ the agreement with the CFT predictions for $k=5$, i.e. $\nu=7/10$ and $\beta=3/20$, is spectacular. This strongly suggests that the continuous transition terminates on both sides at end points in the WZW SU(2)$_5$ universality class. For the two cuts in the central part of the critical line, the values of the critical exponents agree within $5\%$ with the CFT predictions for SU(2)$_3$, $\beta=1/4$ and $\nu=5/6\approx0.833$. For these two cuts one can spot significant finite-size effects that we believe are due to the renormalization to the lower $k=3$ level. We expect crossover effects to be even larger in the vicinity of the end points, resulting in the apparent scaling dimension shown in Fig.\ref{fig:dimeriscaling}(b) that takes intermediate values. Nevertheless, our results provide strong evidence in favor of an extended WZW SU(2)$_3$ line terminated at two WZW SU(2)$_5$ end points as shown in Fig.\ref{fig:phasediag}.

To the best of our knowledge the realization of the WZW SU(2)$_{k=5}$ in a non integrable spin-5/2 chain is reported for the first time. For this reason we perform an additional check of the universality class at the upper end point by looking at the conformal tower of states. According to conformal field theory\cite{diFrancesco}, at the critical point and in the case of conformally invariant boundary conditions, the energy excitation spectrum scales linearly with the system size with a specific tower structure characteristic of the underlying theory.  In our simulations we will only probe the lowest level for each magnetization sector, i.e. the envelop of the tower. The ground-state of the system with $N$-even is a singlet and thus has total spin $j=0$. It appears as a lowest-energy state in the sector wiht zero total magnetization $S^z_\mathrm{tot}=0$. The triplet excitation appears as the lowest-energy state in the sector $S^z_\mathrm{tot}=1$, the quintuplet in the sector $S^z_\mathrm{tot}=2$ etc. We restrict ourselves to $S^z_\mathrm{tot}=7$. For the chain with an odd number of sites, we expect the ground-state to have total spin-5/2 and thus to be described by the $j=5/2$ tower. In appendix \ref{sec:ap1} we construct the envelop of these two towers closely following Ref.\cite{AffleckGepner}. The predictions for the envelops are summarized in Table~\ref{tb:wzwsu2k5}.

\begin{table}[h!]
\centering
\begin{tabular}{|c|c|c|c|c|c|c|c|c|}
\multicolumn{7}{c}{j=0}\\
\hline 
$S^z_\mathrm{tot}$&0&1&2&3&4&5&6&7\\
\hline 
$(E-E_0)N/ \pi v$ &0&1&2&3&4&5&8&11\\
\hline 
\multicolumn{7}{c}{j=5/2}\\
\hline 
$S^z_\mathrm{tot}$&2.5&3.5&4.5&5.5&6.5&7.5&8.5&9.5\\
\hline 
$(E-E_0)N/ \pi v$, &0&2&4&6&8&10&14&18\\
\hline 
\end{tabular}
\caption{Lowest excitation energy with spin $S^z_\mathrm{tot}$ for both $j=0$ and $j=5/2$ WZW SU(2)$_{5}$ conformal towers. See Appendix \ref{sec:ap1} for the construction}
\label{tb:wzwsu2k5}
\end{table}

\begin{figure}[h!]
\centering 
\includegraphics[width=0.49\textwidth]{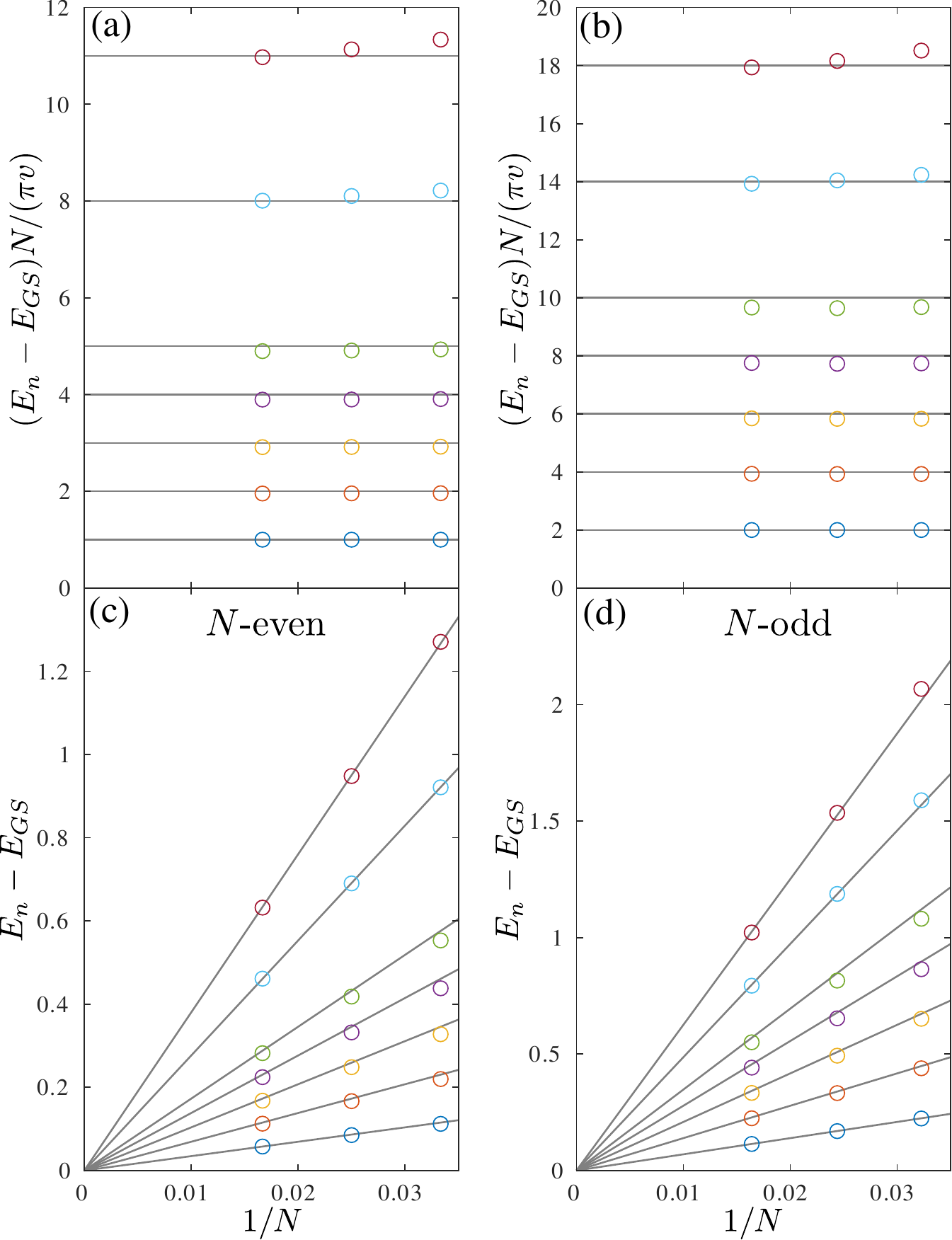}
\caption{ Conformal towers of states (a-b) and finite-size scaling of the excitation spectrum (c-d) at the WZW SU(2)$_5$ end point at $J_2=0.09$, $J_3\approx 0.02325$. Symbols are DMRG data, gray lines are CFT predictions.  
 For $N$-even the ground-state is in the singlet sector computed within $S^z_\mathrm{tot}=0$ ($j=0$ tower), for $N$-odd the ground-state has $S^z_\mathrm{tot}=5/2$  ($j=5/2$ tower).  The towers show only the lowest state within each sector of total magnetization up to $S^z_\mathrm{tot}=7$ for $N$-even and up to $S^z_\mathrm{tot}=9.5$ for $N$-odd.  The non-universal value of the sound velocity $v\approx 3.51$ has been obtained by fitting the singlet-triplet gap for $N$-even with $E_1-E_0\propto\pi v/N$. }
\label{fig:towers}
\end{figure}

In order to extract the tower of states from the energy excitation spectrum one has to know the sound velocity, a non-universal constant. Here we extracted it by fitting the lowest excitation level with $N$-even, building on the fact that the singlet-triplet gap scales as $E_1-E_0\propto\pi v/N$, as shown in Fig.\ref{fig:towers}(c), and we obtained the value $v\approx 3.51$. For all other levels we fix the velocity to this value. The resulting towers are presented in Fig.\ref{fig:towers} and are in excellent agreement with the predicted envelop.

Let us come back to the point postponed in the beginning of this section, namely why, on the commensurate critical side of the transition, the dimerization in a log-log scale does not scale linearly but builds a concave curve. As we now know, the transition, if it is continuous, is either SU(2)$_{k=3}$ or SU(2)$_{k=5}$ at the two end points. In both cases, the scaling dimension at the transition is significantly smaller than the scaling dimension $d=1/2$ of the commensurate critical phase. This implies that, on a finite-size chain, we will see a crossover from  $k=3$ or $k=5$ to $k=1$, and the closer to the transition we are, the larger the system size it will take before the 1/2 slope will appear.

\begin{figure}[h!]
\centering 
\includegraphics[width=0.49\textwidth]{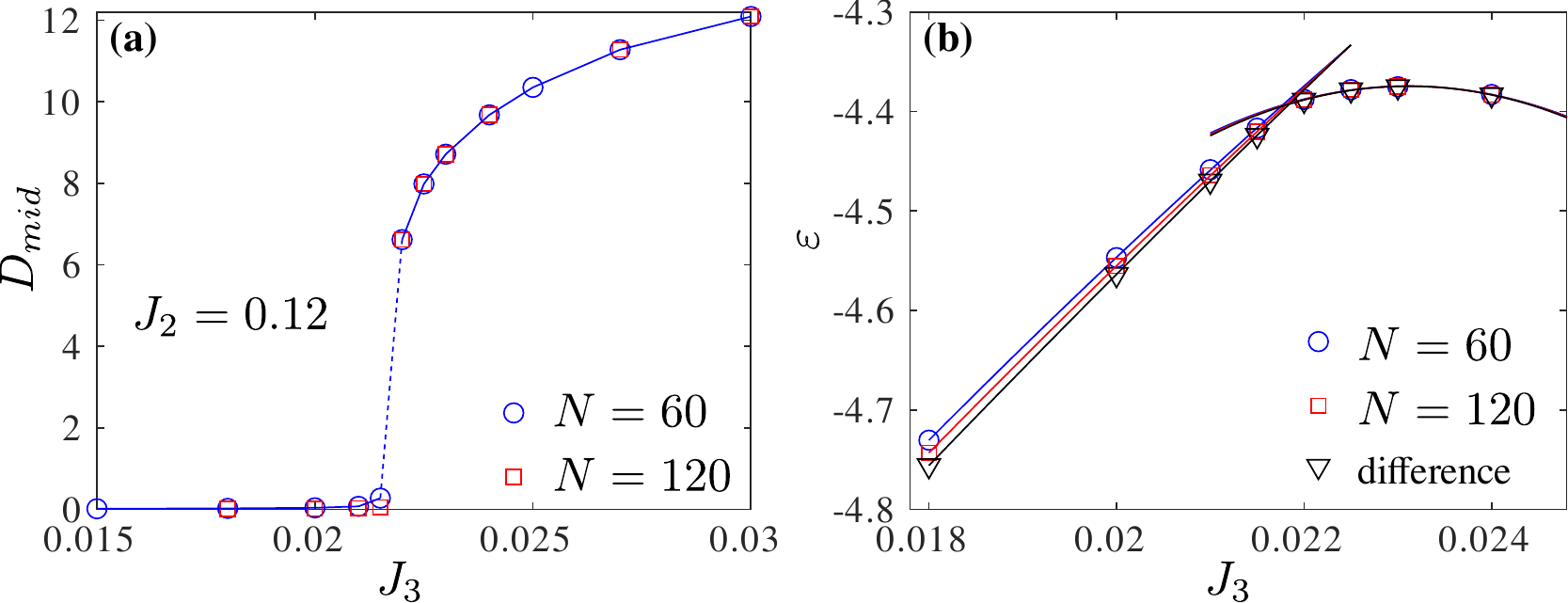}
\caption{ (a) Finite jump in the dimerization and (b) kink in the ground-state energy per site $\varepsilon$ at the first order transition between the commensurate critical and the fully dimerized phase above the upper end point. The results have been obtained for $N=60$ (blue circles) and $N=120$ (red squares). The black triangles stand for $\varepsilon=(E_{120}-E_{60})/60$ - our estimate of the energy per site in the thermodynamic limit. }
\label{fig:1storder}
\end{figure}

Finally, there is still one thing we have not discussed so far: What happens beyond the SU(2)$_5$ end points? The fast decay of the scaling dimension $d$ presented in Fig.\ref{fig:dimeriscaling}(b) excludes the possibility of a continuous transition either $k=1$ or $k=3$ and suggests that the transition becomes first order. For negative values of $J_2$ this is in fact very natural. Indeed, the transition line turns toward the right, and at $J_2\approx-0.25$ it crosses the line defined by Eq.\ref{eq:exact} where the ground-state is given by the exactly dimerized state. This implies that the dimerization jumps from $D=S(S+1)=8.75$ to a value that vanishes algebraically with the system size in the commensurate critical phase, in agreement with a first-order transition. For positive values of $J_2$, the exact line hits the transition when the system is already in the partially dimerized phase, so we cannot rely on the same argument. There we confirm the first-order nature of the transition by the finite-jump in the middle-chain dimerization presented in Fig.\ref{fig:1storder}(a) and by the kink in the ground-state energy per site shown in Fig.\ref{fig:1storder}(b).  In order to eliminate the edge effects on the latter, we also estimated the ground-state energy per site from the difference between the two finite-size values as $\varepsilon=(E_{N=120}-E_{N=60})/60$. The transition we observe at $J_2=0.12$ is very weakly first order and neither the jump in the dimerization nor the kink in the ground-state energy are pronounced. However, since we excluded other WZW transitions, a first-order transition is the only option.

\begin{figure}[t!]
\centering 
\includegraphics[width=0.5\textwidth]{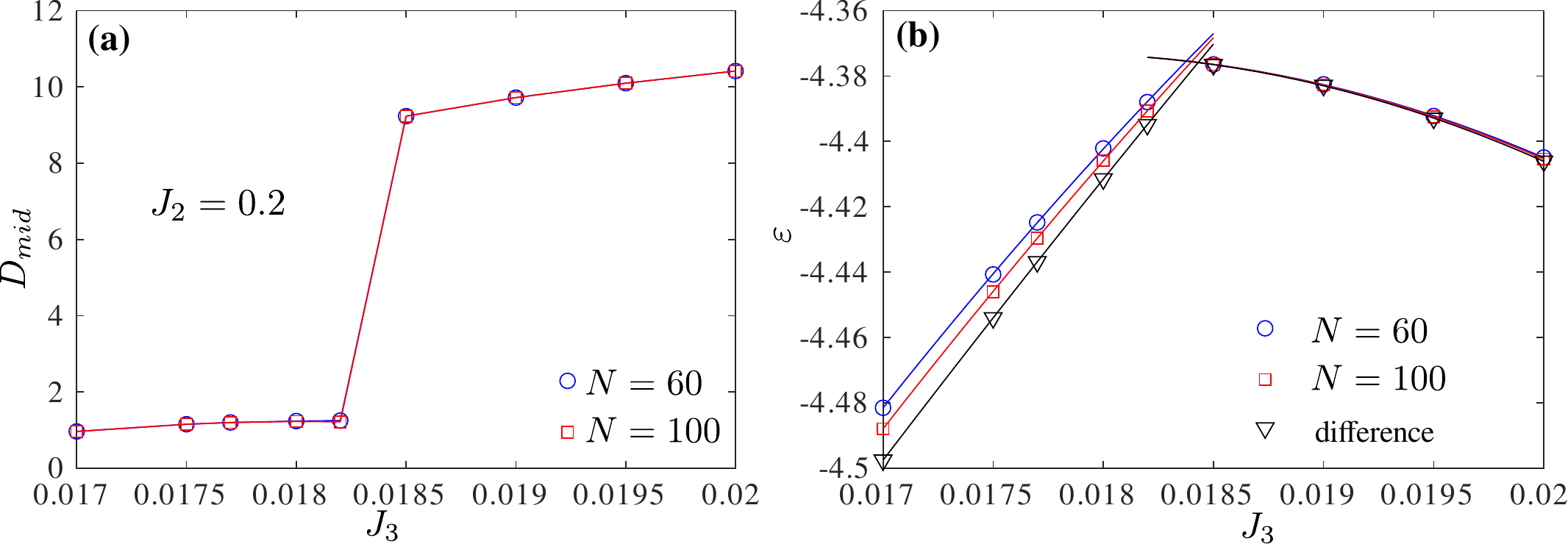}
\caption{ (a) Jump in the middle-chain dimerization and (b) kink in the ground-state energy per site at the first order transition between the partially- and fully-dimerized phases. }
\label{fig:1storder2}
\end{figure}

If we follow the line further up, the first order transition, now between the two dimerized phases, becomes much more pronounced, as shown in Fig.\ref{fig:1storder2}.

\section{Floating phase}
\label{sec:float}

\begin{figure}[h!] 
\centering 
\includegraphics[width=0.49\textwidth]{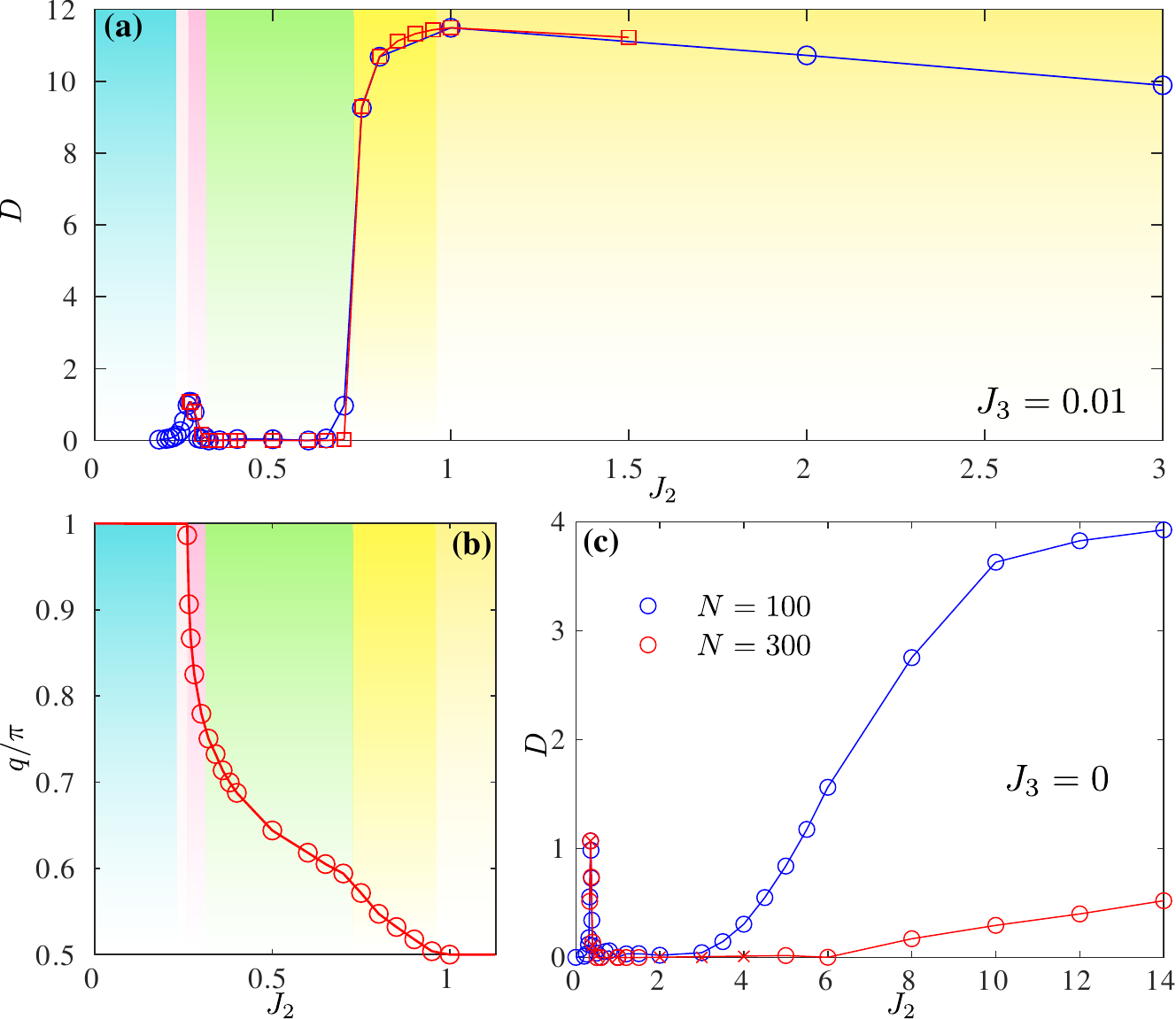}
\caption{ (a) Dimerization and (b) incommensurate wave-vector $q$ along a vertical cut at $J_3=0.01$ as a function of $J_2$. Blue and red symbols correspond to the results obtained with $N=100$ and $N=300$ respectively.  The background colors mark different phases and match those of the main phase diagram in Fig.\ref{fig:phasediag}. (c) Dimerization in the $J_1-J_2$ model ($J_3=0$) as a function of $J_2$. At very large $J_2$, the system enters the second dimerized phase. }
\label{fig:float}
\end{figure}

Let us now focus on another exotic critical regime realized in this phase diagram - the floating phase. According to the classification introduced by Bak\cite{Bak_1982}, the floating phase is a critical phase in which the dominant wave-vector $q$ is not "frozen" to any specific value, but "floats" or changes continuously all over the phase. Let us first show that there is a second critical phase on the phase diagram. For this, we look at the dimerization as a function of the next-nearest-neighbor coupling $J_2$ along a vertical cut at $J_3=0.01$ (see Fig.\ref{fig:float}). For small values of $J_2$ ,the system is in the commensurate critical phase, and the dimerization vanishes. Then, one can clearly distinguish two phases with finite dimerization: a narrow partially-dimerized phase with a maximal dimerization $D\approx 1$ around $J_2\approx 0.3$, and an extended fully dimerized phase above $J_2\approx0.75$. However, between these two dimerized phases there is a finite interval with a vanishing dimerization which, for a half-integer spin chain, implies that the system is gapless according to the Lieb-Schulz-Mattis theorem, hence critical in 1D.

Remarkably enough, all four phases are realized in the pure $J_1-J_2$ model. Our numerical results along the $J_3=0$ axis are presented in Fig.\ref{fig:float}(c). They suggest that the system only enters the second dimerized phase at very large $J_2$. The finite-size effects are very strong, so we cannot determine the critical point or the value of the dimerization with quantitative accuracy. However, thanks to field theory arguments, one knows that, in the limit $J_1/J_2\ll 1$, the interchain zigzag coupling is a relevant perturbation that is expected to induce dimerization, so we are confident that the development of dimerization at large $J_2$ is not a finite-size effect.

\begin{figure}[h!]
\centering 
\includegraphics[width=0.49\textwidth]{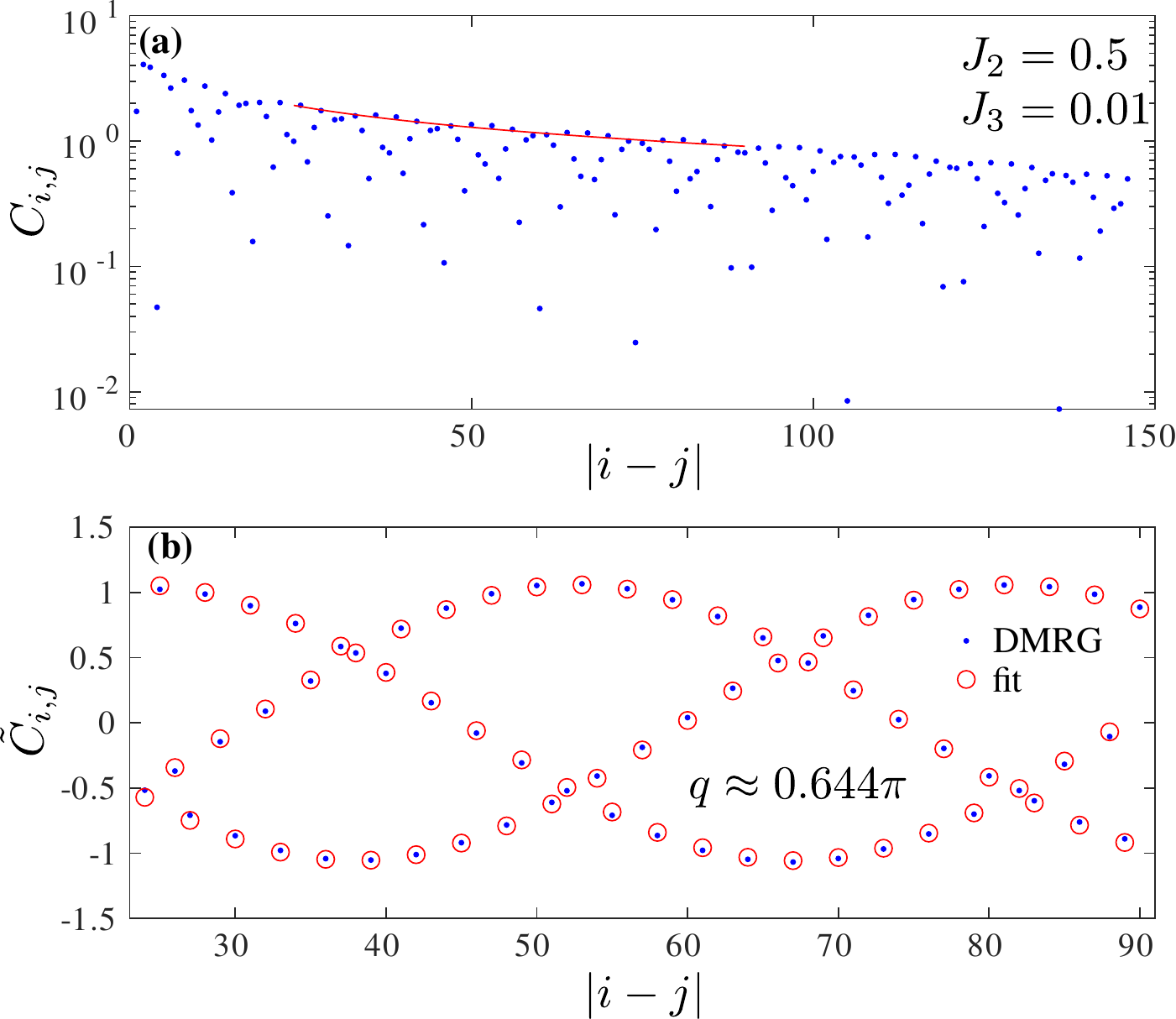}
\caption{ Example of the fit of the spin-spin correlation function at the point inside the floating phase. In the first step (a), we fit the main slope discarding the oscillations. In the second
step (b), we fit the reduced correlation function to extract the wave-vector $q$}
\label{fig:fitexample}
\end{figure}

In order to extract the wave-vector $q$, we fit the spin-spin correlation function $C_{i,j}=\langle {\bf S}_i \cdot {\bf S}_j\rangle-\langle {\bf S}_i\rangle \cdot\langle {\bf S}_j\rangle$ to the Ornstein-Zernicke form:
\begin{equation}
C_{i,j}\propto \frac{e^{-|i-j|/\xi}}{\sqrt{|i-j|}}\cos(q|i-j|+\varphi_0),
\end{equation}
where the effective correlation length $\xi$, the wave vector $q$ and the initial phase $\varphi_0$ are fitting parameters. In the critical phase the correlations are expected to diverge algebraically, however, the finite length of the chain together with the finite bond dimension of the DMRG tensors result in an effective finite correlation length. This allows us to define a unified protocol for both gapped and gapless phases. 
In order to extract the wave-vector with a sufficiently high precision we fit the correlation function in two steps.
 First, we discard the oscillations and fit the main slope of the decay as shown in Fig.\ref{fig:fitexample}(a). This allows us to perform a fit  in a semi-log scale $\log C(x=|i-j|)\approx c-x/\xi-\log(x)/2$. Second we define a reduced correlation function
\begin{equation}
\tilde{C}_{i,j}=C_{i,j} \frac{\sqrt{|i-j|}}{e^{-|i-j|/\xi+c}}
\end{equation}
and fit it with a cosine $\tilde{C}_{i,j}\approx a\cos(q|i-j|+\varphi_0) $ as shown in Fig.\ref{fig:fitexample}(b). The agreement between the DMRG data (blue dots) and the fit (red circles) is almost perfect.

Our results for the dominant wave-vector $q$ along $J_3=0.01$ are summarized in Fig.\ref{fig:float}(b). The system becomes incommensurate inside the partially dimerized phase, remains incommensurate all over the critical (floating) phase, and shortly after the transition to the fully dimerized phase it becomes commensurate again, but with a wave-vector $q=\pi/2$. The evolution of the incommensurate wave-vector all over the floating phase has been indicated in the main phase diagram in Fig.\ref{fig:phasediag} with equal-q lines drawn with a step of $0.05\pi$.

Let us briefly comment on the nature of the phase transitions to the floating phase. The entire floating phase is surrounded by portions of the gapped phases with incommensurate short-range order, and the equal-q lines are continuous through the phase boundary. Therefore, the transitions cannot be classified as commensurate-incommensurate. By analogy with the spin-3/2 case\cite{spin_32paper}, we expect the transition to the partially dimerized phase to be in the Kosterlitz-Thouless universality class. The numerical verification of this prediction is computationally very expensive however, and it goes beyond the scope of this paper. The transition between the floating phase and the dimerized phase has been located by looking at the separatrix in the finite-size scaling of the local dimerization. For the parameter range $J_2\leq 1.5$ of Fig.\ref{fig:phasediag}, the slope of the separartrix, i.e. the effective scaling dimension, never exceeds the value $d\approx 0.2$, which excludes a WZW SU(2)$_k$ transition for all allowed level indices $k=1,3,5$. We therefore conclude that the transition must be first order. This is indirectly confirmed by the kinks in the equal-$q$ lines at this transition, as can be seen in the main phase diagram of Fig.\ref{fig:phasediag} and more clearly in Fig.\ref{fig:float}(b).

A similar magnetic floating phase has been reported recently in the spin-3/2 version of the model\cite{spin_32paper}. It is already well established that in the presence of next-nearest-neighbor interaction the spin-S chain exhibits a disorder point -  a point beyond which the dominant wave-vector becomes incommensurate. When such an incommensurability appears inside a gapped phase, for instance in the Haldane phase of the spin-1 chain, or in the partially-dimerized phases of the spin-3/2 and spin-5/2 chains, the incommensurability only affects the short-range correlations. If, however, the correlations are incommensurate inside a critical phase, the incommensurate order naturally becomes quasi-long-ranged and a floating phase appears.

\section{Discussion}
\label{sec:discussion}

To summarize, the spin-5/2 $J_1-J_2-J_3$ chain leads to a rich phase diagram that contains two dimerized phases and two critical phases.
The number of phases and their relative location are very similar to those reported for spin-3/2 chain\cite{spin_32paper}. This suggests that one can expect a similar phase diagram for all half-integer spin chains with $S\geq 3/2$.  

For spin-5/2,  we have nevertheless reported a number of unexpected features. In particular, we have found the transition between the critical and the fully dimerized phases to be {\it i)} continuous WZW SU(2)$_{k=3}$ for $-0.177\simeq J_2\simeq 0.09$; {\it ii)} WZW SU(2)$_{k=5}$ at both ends of this interval; {\it iii)} first order otherwise. This result agrees with two strong constraints imposed by field theory, namely (i) that the WZW SU(2)$_{k=5}$ universality class in a two-dimensional parameter space can only be realized at isolated points due to the presence of two relevant operators in the theory, and (ii) that the renormalization group flow is only possible between WZW  SU(2)$_{k}$ theories with different level indices $k$ if the parity of the level indices does not change. 

For larger half-integer values of the spin, we expect the same universality classes to be realized if the transition between the SU(2)$_1$ commensurate phase and the dimerized phase is continuous, namely SU(2)$_{k=3}$ along the line, and SU(2)$_{k=5}$ at the ends, because in a two-dimensional parameter space these are the only theories with $k$ odd and $k>1$ that can be stabilized by adjusting 1 and 2 parameters respectively. SU(2)$_k$ theories with $k>5$ have more relevant operators and require more parameters to be fine tuned. 

Moreover, we have reported the emergence of an extended floating phase - an exotic critical phase with quasi-long-range incommensurate order. Compared to the spin-3/2 chain, the floating phase in the spin-5/2 chain is much larger. According to our results, upon increasing $J_2$ to very large values, the system eventually falls into the dimerized phase even for $J_3=0$, but the floating phase extends at least up to $J_2/J_3\sim 6$. For the parameter range shown in Fig.\ref{fig:phasediag}, the transition between the floating phase and the fully-dimerized phase is always first order. However, we cannot exclude that for larger values of $J_2$ the transition will eventually turn into a continuous one. In this case, the most likely scenario is similar to the one realized at small $J_2$ - the transition becomes continuous through an end point in the WZW SU(2)$_5$ universality class and is in the WZW SU(2)$_3$ universality class beyond it.

The floating phase appears to be a generic feature of zig-zag chain with half-integer spins $S\geq 3/2$, and one can thus hope to be able to observe it experimentally. In that respect, the fact that it occupies a significant portion of the phase diagram for $S=5/2$ increases its chances to be realized in experiments. A promising candidate is an iron oxide discovered recently, ${\mathrm{Bi}}_{3}{\mathrm{FeMo}}_{2}{\mathrm{O}}_{12}$. It consists of well separated spin-5/2 chains, with coupling constants $J_2/J_1\approx 1.1$\cite{PhysRevB.104.184402}. The experimental study reports gapless excitations\cite{PhysRevB.104.184402}. According to our theory, given its ratio $J_2/J_1$, this system cannot be in the small $J_2$ commensurate phase. It has to be in the floating phase with incommensurate correlations. It would be very interesting to check experimentally for the presence of this quasi-long-range incommensurate order in this compound.\\

\section{Acknowledgments}
We are indebted to Philippe Lecheminant for pointing out the inconsistency between the prediction of a generic WZW SU(2)$_{2S}$ transition and the presence of two relevant operators in the WZW SU(2)$_5$ theory. We also thank Ciaran Hickey for bringing to our attention the recently discovered iron oxide compound ${\mathrm{Bi}}_{3}{\mathrm{FeMo}}_{2}{\mathrm{O}}_{12}$. 
This research has been supported by Delft Technology Fellowship (NC) and by the Swiss National Science Foundation (FM). The numerical simulations have been performed using the facilities of the Scientific IT and Application Support Center of EPFL and on the Dutch national e-infrastructure with the support of the SURF Cooperative.

\begin{appendix}

\section{Construction of WZW SU(2)$_5$ conformal towers of states}
\label{sec:ap1}

We construct the  conformal towers of states for WZW SU(2)$_{k=5}$ critical chains closely following the procedure introduced in Ref.\cite{AffleckGepner}. Let us remind it here: Starting from any state with some value of $S^z_\mathrm{tot}$ ($S^z_L$ in the original notations) and level $n$, follow the diagonal line to the point $(-k\ \mathrm{sign}(S^z_\mathrm{tot})-S^z_\mathrm{tot},n+k+2|S^z_\mathrm{tot}|)$.
All grid nodes on this line will have a non-zero multiplicity. 

\begin{widetext}

\begin{figure}[t!]
\centering 
\includegraphics[width=0.9\textwidth]{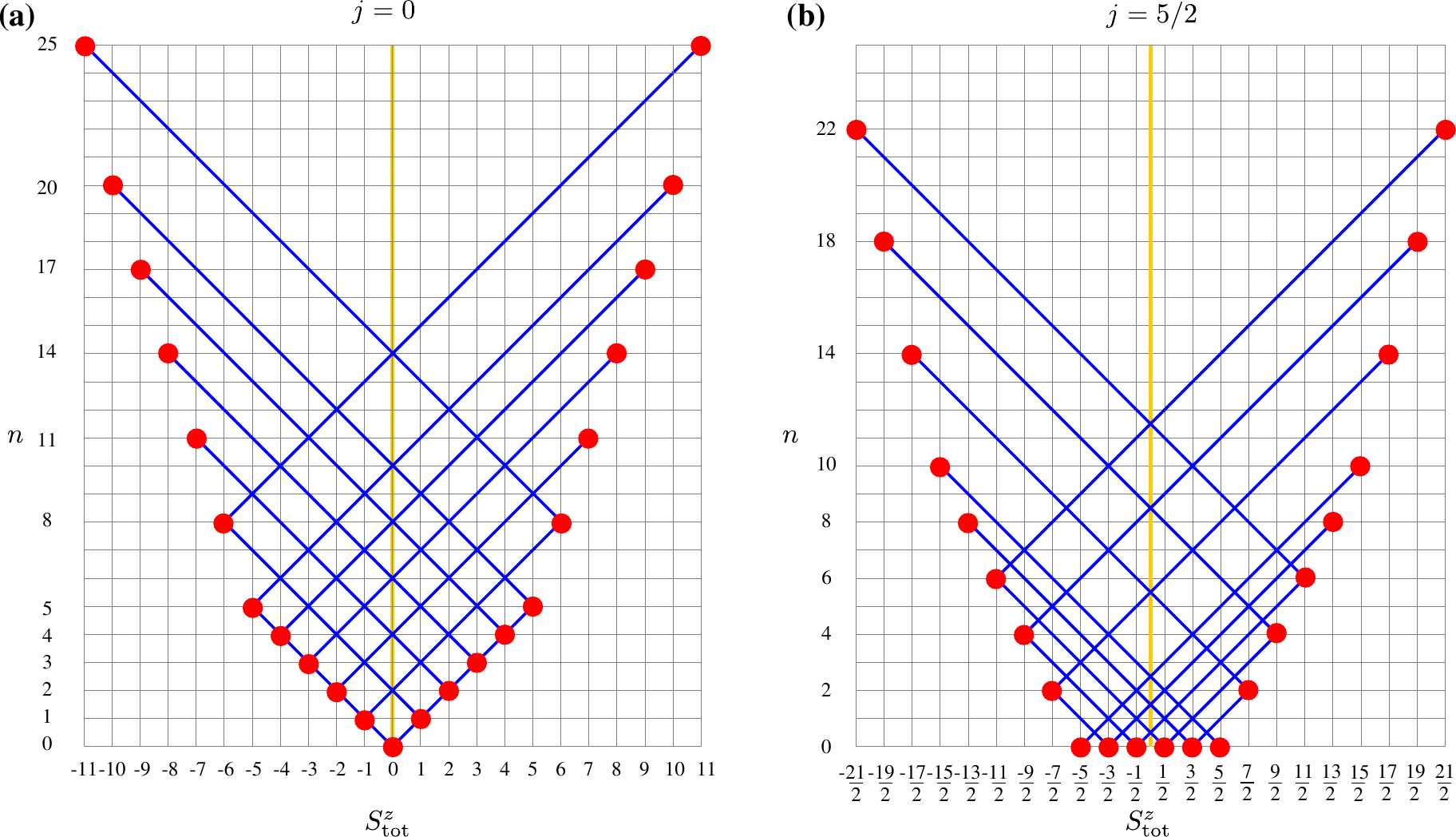}
\caption{ Construction of the conformal towers of states of WZW SU(2)$_{k=5}$ antiferromagnetic chains with total spin {\bf (a)} $j=0$ and {\bf(b)} $j=5/2$. Starting from the ground-state at $n=0$ we follow the diagonal lines (blue) as described in the main text to find the states with non-zero multiplicity. Every such state can serve as a new starting point for another diagonal line. The states that have the lowest $n$ for a fixed $S^z_\mathrm{tot}$ form the tower's envelop and are marked with filled red circles. The middle line of the tower $S^z_\mathrm{tot}=0$ is shown in yellow for convenience.  }
\label{fig:CFTtowers}
\end{figure}

\end{widetext}

For $j=0$ we start with the ground state that has $S^z_\mathrm{tot}=0$. In order to apply the above rule  we assume $S^z_\mathrm{tot}=0+$ or $0-$. This define two lines that start at (0,0) and stop at (5,5) and (-5,5). All states between and including these two points have non-zero multiplicity. So, we can chose any node on this line as a new starting point to continue the construction. The obtained envelop is shown in Fig.\ref{fig:CFTtowers}(a).

For the chain with an odd number of sites we expect the ground-state to have a total spin $j=5/2$. This implies that the ground-state appears as the lowest energy state in {\it all} sectors with total magnetization ranging from $S^z_\mathrm{tot}=-5/2$ to $5/2$. Starting from $n=0$ and $S^z_\mathrm{tot}=\pm 3/2$ and $\pm 1/2$ one can also go to the points $(k\ \mathrm{sign}(S^z_\mathrm{tot})-S^z_\mathrm{tot}, n+k-2|S^z_\mathrm{tot}|)$.


\end{appendix}

\bibliographystyle{apsrev4-2}
\bibliography{bibliography}

\end{document}